\documentclass[twocolumn,aps,amsmath,amssymb,showpacs,floatfix]{revtex4}
\usepackage{graphicx}
\usepackage{dcolumn}
\usepackage{bm}
\usepackage{hyperref} 
\usepackage{latexsym}
\usepackage{float}

\def\av#1{\langle#1\rangle}
\def\df{d_{\text{f}}}

\begin{document}
\title{Kleinberg Navigation in Fractal Small World Networks}   
\author{Mickey R. Roberson}
\author{Daniel ben-Avraham}
\email{benavraham@clarkson.edu}
\affiliation{Department of Physics, Clarkson University,
Potsdam NY 13699-5820}

\begin{abstract} 
 We study the Kleinberg problem of navigation in Small World networks when the underlying lattice is a fractal consisting of $N\gg1$ nodes.  Our extensive numerical simulations confirm the prediction that most efficient navigation is attained when the length $r$ of long-range links is taken from the distribution $P({\bf r})\sim r^{-\alpha}$, where $\alpha=\df$,  the fractal dimension of the underlying lattice.  We find finite-size corrections to the exponent $\alpha$, proportional to  $1/(\ln N)^2$.  
 \end{abstract}

\maketitle

\date{\today}


Recently Kleinberg has studied the problem of efficient navigation in {\it Small World\/} networks, based on local algorithms that rely on the underlying geography~\cite{kleinberg00a,kleinberg00b}.  Consider for example a $d$-dimensional hypercubic lattice, consisting of $N$ nodes, where in addition to the links between nearest neighbors each node $i$ is connected randomly to a node $j$ with a probability proportional to $r_{ij}^{-\alpha}$, (here, and elsewhere, $r_{ij}=|{\bf r}_i-{\bf r}_j|$ denotes the Euclidean distance between nodes $i$ and $j$).  Suppose that a message is to be passed from a ``source" node $s$ to a ``target" node $t$, along the links of the network, by a {\it decentralized\/}, or {\it local\/} algorithm (an algorithm that relies solely on data gathered from the immediate vicinity of the node that holds the message at each step), when the location of the target is publicly available.  Kleinberg shows that when the exponent $\alpha=d$ an algorithm exists that requires less than $(\ln N)^2$ steps to complete the task.  If $\alpha\neq d$, the required number of steps grows as a power of $N$. Moreover, no local algorithm will do better, functionally, than the simple-minded {\it greedy\/} algorithm:
{\it pass the message forward to the neighbor node which is closest to the target\/} (geographically).

Kleinberg observes~\cite{kleinberg00b} that the above results generalize to ``less structured graphs with analogous scaling properties."  Interest in such cases is practical, as the nodes of many real-life networks (routers of the Internet, population in social nets, etc.) are not distributed regularly.  Here we test this assertion for the case of {\it fractal} lattices, enhanced by the addition of random long-range links as in the original Kleinberg problem.  We find that
the results indeed generalize to this case, and that most efficient navigation is achieved when the
power exponent for the random connections is $\alpha=\df$, the fractal dimension of the underlying lattice.   Our numerical analysis is sensitive enough to allow for a study of finite-size effects.  For a lattice of $N$ nodes optimal navigation is attained for an {\it effective\/} exponent $\alpha(N)$ that is smaller than the idealized limit of $\alpha=\df$ (when $N\to\infty$) by as much as $1/(\ln N)^2$.
Thus, corrections are substantial even for very large lattices.

\begin{figure}[ht]
  \vspace*{0.cm}\includegraphics*[width=0.36\textwidth]{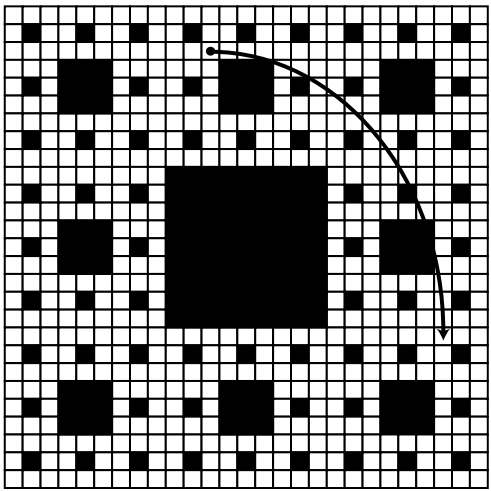}
\caption{Small World net based on the Sierpinski carpet.  Shown is a carpet of generation $n=3$.  The nodes (empty squares) are connected to their nearest neighbors (not shown).  In addition, each node $i$
is connected to a random node $j$ as described in the text.  For the sake of clarity, only one such connection is shown as an example.}
\label{carpet}
\end{figure}

Consider a fractal  lattice, such as the Sierpinski carpet~\cite{mandelbrot} (Fig.~\ref{carpet}), where, in addition to the existing links, each node $i$ is randomly connected to a single node $j$, selected from among all other nodes with probability
$p_{ij}=r_{ij}^{-\alpha}/\sum_k r_{ik}^{-\alpha}$.  The sum in the denominator runs over all nodes $k\neq i$
and is required for normalization.  If the fractal is finite, consisting of $N\gg1$ nodes, its {\it linear\/} size is 
$L\sim N^{1/\df}$.  The normalization term then scales as
\begin{equation}
\label{normalize}
\sum_k r_{ik}^{-\alpha}\sim\int_1^L r^{-\alpha}\,r^{\df-1}dr\sim
\begin{cases}
(\alpha-\df)^{-1} &\alpha>\df,\\
\ln L &\alpha=\df,\\
L^{\df-\alpha} &\alpha<\df.
\end{cases}
\end{equation}

The average distance between randomly chosen $(s,t)$-pairs is $\sim L$, so in the absence of long contacts a message takes  $T\sim N^{1/\df}$ steps to be delivered~\cite{remark}.  Long range contacts reduce the $1/\df$ exponent, but only when the exponent $\alpha=\df$ does the expected delivery time scale 
slower than a power of $N$.
The basic idea of Kleinberg's argument, applied to the case of fractals, is as 
follows~\cite{kleinberg00b}.  For $\alpha=\df$, surround the target node $t$ with $m$-shells of radii
$e^{m-1}<r<e^m$, $m=1,2,\dots$ Suppose that the message holder is currently in shell $m$, then the probability that the node is connected by a long-range link to a node in shell $m-1$, is, according to~(\ref{normalize}),
\[
\sim\int_{e^{m-1}}^{e^{m}}\frac{r^{-\df}}{\ln L}\,r^{\df-1}dr=\frac{1}{\ln L}\;.
\]
The probability to reach the next shell ($m-1$) in more than $x$ steps is then $\psi(x)=(1-1/\ln L)^x$, and the average number of steps required to do so is~\cite{derive} 
\[
\av{x}=\int_0^{\infty}\psi(x)\,dx\sim\ln L\;.
\]
Since the largest shell is for $e^m=L$, the number of shells between the source and target is 
of the order of $m=\ln L$.  Thus, the expected total number of steps required to reach the target 
is $\sim(\ln L)^2$~\cite{remark1}.

For $0<\alpha<\df$, surround the target node $t$ by a ball of radius $\ell=L^{\delta}$, ($0<\delta<1$).  The probability that a randomly chosen node, $i$, has a long range contact to a site $j$ within the ball, is, according to~(\ref{normalize}), $\sim r_{ij}^{-\alpha}/L^{\df-\alpha}\leq 1/L^{\df-\alpha}$. Thus, the probability that $i$ connects to {\it any\/} node in the ball does not exceed 
$\ell^{\df}/L^{\df-\alpha}=L^{\delta\df-\df+\alpha}$.  Since $\delta<1$, the source lies almost surely outside of the ball (in the limit $N\to\infty$).  Then, any $\ell$-step path between the source and target must contain at least one long range connection into a node inside the ball.  But the probability that a node with such a connection is encountered within $\ell$ steps is smaller 
than $\ell\times L^{\delta\df-\df+\alpha}$.
If this probability vanishes, as $N\to\infty$, the expected number of steps is bound to be at least $\ell$.  This happens for $\delta<(\df-\alpha)/(1+\df)$, so the expected number of steps exceeds
$L^{(\df-\alpha)/(1+\df)}$.

For $\alpha>\df$, the probability that a node has a long-range connection longer than $L^{\gamma}$,
($0<\gamma<1$) scales, according to~(\ref{normalize}), like
\[
\int_{L^{\gamma}}^{\infty}\frac{r^{-\alpha}}{\alpha-\df}\,r^{\df-1}dr\sim L^{\gamma(\df-\alpha)}\;.
\]
Thus, the probability to jump a distance larger than $L^{\gamma}$ within $L^{\beta}$ steps, ($0<\beta<1$), is less than $L^{\beta}L^{\gamma(\df-\alpha)}$.
If this probability vanishes, as $N\to\infty$, then the total distance covered cannot exceed 
$L^{\beta}L^{\gamma}$.  Since the expected distance between source and target is of order $L$,
we need $\beta+\gamma=1$.  On the other hand, the probability for steps longer than $L^{\gamma}$
vanishes when $\beta+\gamma(\df-\alpha)<0$.  
The two conditions yield $\beta<(\alpha-\df)/(\alpha-\df+1)$, so the expected number of steps exceeds
$L^{\beta}\sim L^{(\alpha-\df)/(\alpha-\df+1)}$.

We have simulated navigation by the greedy algorithm in Small World nets based on the Sierpinski carpet (Fig.~\ref{carpet}).    A finite carpet, constructed recursively to generation $n$, consists of 
$N=8^n$ nodes arranged within a square of side $L=3^n$.  In each run random nodes are selected
as the source ($s$) and target ($t$) and a path is sought between the two by the greedy algorithm.
To minimize computer time and memory, each successive node and its random long-range connection is constructed only as it is reached by the message.  In this way we were able to simulate carpets of up to generation $n=19$ ($L\approx1.2\times10^{9}$, $N\approx1.4\times10^{17}$).
For each value of the exponent $\alpha$, the expected number of steps was obtained from averaging over 10,000 runs, apart from the largest lattice, of generation $n=19$, for which the number of runs per data point was reduced to 1,000.  

\begin{figure}[ht]
  \vspace*{0.cm}\includegraphics*[width=0.40\textwidth]{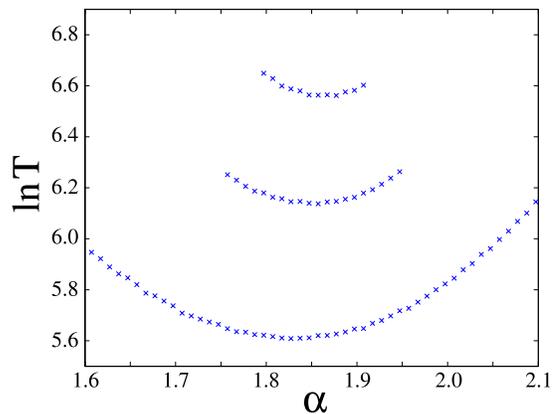}
\caption{(Color online) Delivery time, $T$, as a function of the long-contact exponent, $\alpha$.  Plotted are results for the Sierpinski carpet of generation $n=12$, 15, and 18 (bottom to top). }
\label{TofAlpha}
\end{figure}

An example of the results obtained in this way is shown in Fig.~\ref{TofAlpha}, where we plot the logarithm  of the ``time" $T$ --- the average number of steps required by the greedy algorithm --- as a function of $\alpha$, for nets of generation $n=12$, 15, and 18.  Notice the parabolic shape of the curves, similar to what is observed for the regular square lattice~\cite{kleinberg00a}, as well as the increase in $T$ as the lattice grows larger.  

\begin{figure}[ht]
  \vspace*{0.cm}\includegraphics*[width=0.40\textwidth]{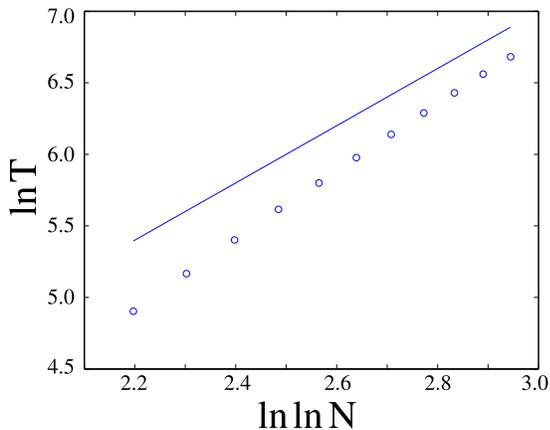}
\caption{(Color online) Delivery time as a function of size.  The logarithm of $T$ is plotted against $\ln\ln N$.  The data
points are for carpets of generation $n=9,10,\dots,19$ (from left to right).  The straight line of slope 2, corresponding to the theoretical prediction $T\sim(\ln N)^2$, is shown for comparison. }
\label{TofN}
\end{figure}

To extract further information, we have fitted second-order polynomials to the $\ln T(\alpha)$ curves of
Fig.~\ref{TofAlpha}.  The fits allow us to compute $\alpha(N)$; the location of the minimum (or the optimal long-contact exponent) for a lattice of $N$ nodes, as well as $T(N)$; the minimal time required, on average, to deliver the message in a lattice of size $N$.  In Fig.~\ref{TofN} we show the dependence of $T(N)$ upon the lattice size.  While it is quite hopeless to study the scaling $T(N)\sim(\ln N)^2$ numerically (since it requires a double logarithmic function of $N$), the slope of $\ln T$ versus $\ln\ln N$ is seen to vary slowly and plausibly converging to $2$, as predicted by Kleinberg. 

\begin{figure}[ht]
  \vspace*{0.cm}\includegraphics*[width=0.40\textwidth]{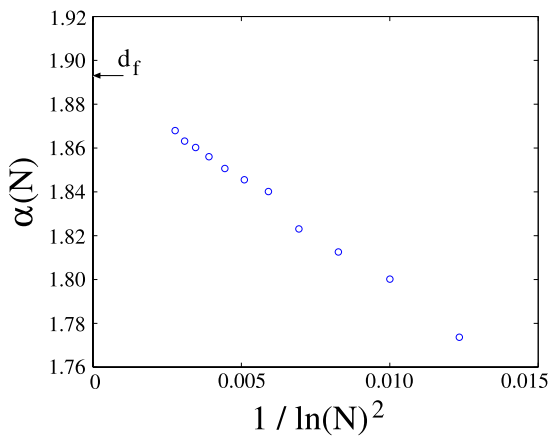}
\caption{(Color online) Optimal long-contact exponent.  The exponent $\alpha(N)$, required for optimal navigation in a finite network of $N$ nodes, is plotted against $1/(\ln N)^2$, for carpets of generation $n=9,10,\dots,19$ (right to left).  The arrow pointing at the expected limit of $\alpha(\infty)=\df=\ln8/\ln3$ is shown to guide the eye.}
\label{AlphaofN}
\end{figure}

In Fig.~\ref{AlphaofN} we show the dependence of $\alpha(N)$ upon $N$.  The optimal value of the exponent for long contacts varies very slowly with the size of the lattice.  Inspired by the scaling $T(N)\sim(\ln N)^2$, we guess 
\begin{equation}
\label{corrections}
\alpha(N)\sim\alpha(\infty)+A/(\ln N)^2\;,
\end{equation}
where $A$ is a constant.
This functional form, as well as the predicted limit of $\alpha(\infty)=\lim_{N\to\infty}\alpha(N)=\df$, for the case of an infinite lattice, is well supported by the data.  
The broadness of the $\ln T(\alpha)$ curves (Fig.~\ref{TofAlpha}) makes it difficult to pinpoint their minima,
$\alpha(N)$, leading to the large fluctuations evident in the plot.  (On the other hand, $T(N)$ can be determined quite accurately from the data, see Fig.~\ref{TofN}.)

In conclusion, we have studied navigation by the greedy algorithm on fractal Small World networks with random long-range connections taken from a power-law distribution.  Our numerical results support the prediction of Kleinberg that optimal navigation occurs when the long-contact exponent is $\alpha=\df$, the fractal dimension of the underlying lattice.  

An important effect are the corrections due to the finite size of the networks involved, expressed in Eq.~(\ref{corrections}).  Even for our largest net, of $N\approx1.4\times10^{17}$ nodes, the correction to $\alpha$
is as big as $1.5\,\%$.  More importantly, the correction is huge for sizes relevant to every-day life.  For nets of
$10^7$ nodes (comparable to the size of the Internet, say), the correction is
nearly $10\,\%$~\cite{remark2}.  It remains an open question whether corrections scale
as $1/(\ln N)^2$ generically, as observed in our case.

We have also carried out simulations on a Small World network based on the Sierpinski gasket~\cite{mandelbrot}.  The results are similar to the ones shown here for the carpet, however, we observe a small but persistent discrepancy between the extrapolated $\alpha(\infty)\approx1.573$ and $\df=\ln3/\ln2\approx1.585$ of the Sierpinski gasket.  For the ease of programming, we had embedded the gasket in a {\it square\/} lattice, thus distorting its original shape (of an equilateral triangle) to a right angled triangle.  This introduces an anisotropy in the distribution of long-contact links: connections along the stretched direction are less
favorable.  We suspect that this anisotropy is the source of the discrepancy.  Anisotropy effects will be considered in detail in future work.

\acknowledgments

We thank H. Rozenfeld, J. Bagrow, and E. Bollt for numerous useful discussions.  We gratefully acknowledge NSF awards PHY0140094 and PHY0555312 (DbA) for partial support of this work.

 

\begin{thebibliography}{99}

\bibitem{kleinberg00a}
J. M. Kleinberg,
Nature {\bf 406}, 845 (2000).

\bibitem{kleinberg00b} 
J. M. Kleinberg,
Conf. Proc. Annu. ACM Symp. Theory Comput., pp.~163-170, 2000 (2000).

\bibitem{mandelbrot} 
B. B. Mandelbrot, {\it Fractals: Form, Chance and Dimension\/}, (Freeman, San Francisco, 1977).

\bibitem{remark} We assume that the chemical distance (the shortest number of links connecting two nodes) is proportional to the Euclidean distance.  We also assume that the fractal in question does not have ``overhangs" --- nodes where the message gets stuck, unless one passes it backwards.  If the typical size of overhangs does not increase with lattice size, an aptly modified greedy algorithm achieves the same results stated here.  The situation is unclear when overhangs grow with lattice size (as in the incipient infinite percolation cluster, say).

\bibitem{derive} Let $p(x)dx$ be the probability to reach the next shell in between $x$ and $x+dx$ steps.  Then, $\psi(x)=\int_x^{\infty}p(x)\,dx$, while $\av{x}=\int_0^{\infty}xp(x)\,dx$.  Integration by parts
of the last relation, in the limit of $L\to\infty$, yields the result in the text.

\bibitem{remark1}
With a little more care one could show that $(\ln N)^2$ is an {\it upper bound\/} for the expected number
of steps, in a similar way as Kleinberg does for the two-dimensional square lattice~\cite{kleinberg00b}.  Likewise,
the expected number of steps cited for $\alpha\neq\df$ can be shown to be {\it lower bounds\/}, but we omit rigor, for simplicity.

\bibitem{remark2} A correction of similar size can be seen in \cite{kleinberg00a}, for a square lattice of $20,000\times20,000$ nodes.

\end{thebibliography}
\end{document}